\documentstyle[12pt,epsf]{article}
\renewcommand{\bar}[1]{\overline{#1}}

\renewcommand{\d}{{\mathrm d}}

\begin{document}

\begin{flushright}
USM-TH-129 
\end{flushright}
\bigskip\bigskip


\centerline{\large \bf Reanalysis of Azimuthal Spin Asymmetries of
Meson Electroproduction}

\vspace{18pt} \centerline{\bf Bo-Qiang Ma
$^{a}$,
Ivan Schmidt
$^{b}$,
and Jian-Jun Yang
$^{b,c}$}

\vspace{8pt}

{\centerline {$^{a}$Department of Physics, Peking University,
Beijing 100871, China,}}

{\centerline {and CCAST (World Laboratory),
P.O.~Box 8730, Beijing 100080, China}}


{\centerline {$^{b}$Departamento de F\'\i sica, Universidad
T\'ecnica Federico Santa Mar\'\i a,}}

{\centerline {Casilla 110-V, 
Valpara\'\i so, Chile}}

{\centerline {$^{c}$Department of Physics, Nanjing Normal
University, Nanjing 210097, China}}

\vspace{6pt}
\begin{center} {\large \bf Abstract}

\end{center}
The azimuthal spin asymmetries for pion production in
semi-inclusive deep inelastic scattering of unpolarized charged
lepton beams on longitudinally polarized nucleon targets, are
reanalyzed by taking into account an important sign correction to
previous formulas. It is found that different approaches of
distribution functions and fragmentation functions may lead to
distinct predictions on the azimuthal asymmetries measured in the
HERMES experiments, thus the available data cannot be considered
as a direct measurement of quark transversity distributions,
although they still can serve to provide useful information on
these distributions and on T-odd fragmentation functions.
Predictions of the azimuthal spin asymmetries for kaon production
are also presented, with different approaches of distribution and
fragmentation functions. The unfavored fragmentation functions
cannot be neglected for $K^-$ and $K^0_S$ production in
semi-inclusive deep inelastic processes.

\vfill

\centerline{PACS numbers: 13.87.Fh, 13.60.-r, 13.88.+e, 14.20.Dh}

\vfill

\newpage


The HERMES collaboration reported the observation of single-spin
azimuthal asymmetries for pion production in semi-inclusive deep
inelastic scattering (DIS) of unpolarized positron beam on the
longitudinally polarized nucleon target \cite{HERMES00,HERMES01}.
Such azimuthal asymmetries are important because they could
provide information on the chiral-odd transversity distributions
and T-odd fragmentation functions, which are less known than the
usual distribution functions and fragmentation functions. The
experimental measurements of quark transversity distributions are
difficulty, since the transversity is not directly observable in
inclusive DIS processes. It has been proposed that the
transversity can manifest itself through the Collins effect
\cite{Col93} of nonzero production between a chiral-odd structure
function and a T-odd fragmentation function, which is accessible
in some specific semi-inclusive hadron production experiments
\cite{Col93,Kot95,Ans95,Mul96,Kot97,Jaf98,Ams}. There have been a
number of studies
\cite{Kot99,Efr00,Sch00,Bor99,Bog00,Ans00,Suz00,MSY9,Oga98,Oga00,Kor00,Efr01,MSY10,EGS}
to show that the azimuthal asymmetries measured by HERMES are
related to the quark transversity distributions of the nucleon.
However, for the azimuthal spin asymmetries of meson production in
DIS processes of unpolarized charged lepton beams on
longitudinally polarized nucleon targets, it just became clear
recently \cite{EGS} that there was an error in the formulas used
in previous calculations
\cite{Efr00,MSY9,Oga98,Oga00,Efr01,MSY10}. It is thus necessary to
reanalyze and check the influences on the calculated results and
conclusions from such a correction. It is the purpose of this note
to update the analysis of our results presented in
Ref.~\cite{MSY10}, and to extend them in order to also predict
azimuthal spin asymmetries for kaon production .

The analyzing power of azimuthal spin asymmetry measured by HERMES
is defined as
\begin{equation}
A_{UL}^W=\frac{\int \left[{\mathrm d} \phi\right] W(\phi)
\left\{N^+(\phi)-N^-(\phi)\right\}}{\frac{1}{2}\int \left[{\mathrm
d} \phi\right]\left\{ N^+(\phi)+N^-(\phi) \right\}}, \label{AP}
\end{equation}
where $UL$ denotes {\it unpolarized} beam on a {\it
longitudinally} polarized target, $W(\phi)=\sin \phi$ or $\sin 2
\phi$ is the weighting function for picking up the Collins effect,
and $N^+(\phi)$ ($N^-(\phi)$) is the number of events for meson
production, as a function of $\phi$, when the target is positively
(negatively) polarized. The azimuthal angle $\phi$ is the angle
between the meson emitting plane and the lepton scattering plane,
with the lepton scattering plane determined by the incident and
scattered leptons, and the meson emitting plane determined by the
final detected meson and the virtual photon. The virtual photon
acts as the common axis of both planes. The analyzing powers of
azimuthal asymmetries for pions have been measured by the HERMES
collaboration \cite{HERMES00,HERMES01}, and there is clear
evidence for non-zero values of $A_{UL}^{\sin \phi}$ for $\pi^+$
and $\pi^0$ production, which indicates the existence of azimuthal
asymmetries.

One can relate the theoretical calculations with the azimuthal
spin asymmetry by
\begin{equation}
A_{UL}^{\sin \phi}=\left< \frac{|P_{h\perp}|}{M_{h}} \sin \phi
\right> =\frac{\Sigma_2}{\Sigma_1}, \nonumber \label{SA1phi}
\end{equation}
where a sum over all quark flavors, $\Sigma_i=\sum_q e_q^2
\Sigma_i^q$, is implicitly assumed, and will be assumed from now
on. In the case of unpolarized beam and longitudinally polarized
target, $\Sigma_1$ and $\Sigma_2=\Sigma_{2L}+\Sigma_{2T}$ are
given \cite{Kot95,Mul96,Oga98,Oga00} by
\begin{equation}
\Sigma_1=\left[1+(1-y)^2\right]f_1(x) D_1(z),
\end{equation}
\begin{equation}
\Sigma_{2L}=4 S_L \frac{M}{Q}(2-y)\sqrt{1-y} \left[ x \, h_L(x) z
H_1^{\perp (1)}(z) - h_{1L}^{\perp (1)}(x) \tilde{H}(z)\right],
\end{equation}
\begin{equation}
\Sigma_{2T}={\mathbf{-}}2 S_{T x}(1-y) h_1(x) z H_1^{\perp (1)}(z)
. \label{Sigma2T}
\end{equation}
We should emphasis here that the ``${\mathbf{-}}$" sign in front
of the right side of (\ref{Sigma2T}) is the correction \cite{EGS}
to Ref.~\cite{MSY10}, and it brings significant changes to the
numerical results presented there. In the above formulas, the
twist-2 distribution functions and fragmentation functions have a
subscript ``1": $f_1$ and $D_1$ are the usual unpolarized
distribution and fragmentation function; $h_{1L}^{\perp (1)}(x)$
and $h_1(x)$ are the quark transverse spin distribution functions
of longitudinally and transversely polarized nucleons,
respectively; $h_L(x)$ is the twist-3 distribution function of a
longitudinally polarized nucleon, and it can be split into a
twist-2 part, $h_{1L}^{\perp (1)}(x)$, and an interaction
dependent part, $\tilde{h}_L(x)$:
\begin{equation}
h_L(x)=-2 \frac{h_{1L}^{\perp (1)}(x)}{x}+\tilde{h}_L(x).
\label{HLL}
\end{equation}
The fragmentation function $\tilde{H}(z)$ is the interaction
dependent part of the twist-3 fragmentation function:
\begin{equation}
H(z)=-2 z H_1^{\perp (1)}(z)+\tilde{H}(z).
\end{equation}
The functions with superscript ``(1)" denote $p_{\perp}^2$- and
$k_{\perp}^2$-moments, respectively:
\begin{equation}
h_{1L}^{\perp (1) }(x) \equiv \int \d^2 p_{\perp}
\frac{p_{\perp}^2}{2M^2}h_{1L}^{\perp }(x,p_{\perp}^2 ),
\end{equation}
\begin{equation}
H_{1L}^{\perp (1) }(z) \equiv  z^2 \int \d^2 k_{\perp}
\frac{k_{\perp}^2}{2M_h^2}H_{1L}^{\perp }(z,z^2k_{\perp}^2 ),
\end{equation}
where $p_{\perp}$ and $k_{\perp}$ are the intrinsic transverse
momenta of the initial and final partons in the target and
produced hadrons, respectively.

To calculate the spin asymmetries and compare them with
experiments, we need the quark distribution functions: $f_1(x)$,
$h_1(x)$, $\tilde{h}_L(x)$, and $h_{1L}^{\perp (1) }(x)$, and the
fragmentation functions: $D_1(z)$, $H_1^{\perp (1)}(z)$, and
$\tilde{H}(z)$. Most of the distribution functions and
fragmentation functions in these expressions are not known a
priori, since they have not been measured yet. Thus we have to
make some assumptions and approximations, and this leads to
different approaches for the distribution functions and
fragmentation functions:


{\it Leading Approach} is to neglect the $1/Q$ term $\Sigma_{2L}$
in $\Sigma_{2}$, i.e., we neglect both the ${\tilde h}_L(x)$ and
$h_{1L}^{\perp (1) }(x)$ terms in the spin asymmetry $\left<
\frac{|P_{h\perp}|}{M_{h}} \sin \phi \right>$. Then we find
immediately that \cite{Efr00,MSY9}
\begin{equation}
\left< \frac{|P_{h\perp}|}{M_{h}} \sin \phi \right>
=\frac{\Sigma_{2T}}{\Sigma_1} \propto {\mathbf{-}}  \frac{h_1(x)
H_1^{\perp (1)}(z) }{f_1(x) D_1(z)}.
\end{equation}
We emphasize here that the ``${\mathbf{-}}$" sign has been added,
and it brings an opposite trend for the azimuthal asymmetries in
comparison with the earlier predictions given in
\cite{Efr00,MSY9}.

{\it Approach 1} is to assume that the twist-2 quark transverse
spin distribution function of longitudinally polarized nucleon,
$h_{1L}^{\perp (1) }(x)$, is zero \cite{Oga00}. Then it follows
that
\begin{equation}
h_L(x)=\tilde{h}_L(x)=h_1(x).
\end{equation}
Notice that in this approach, the spin asymmetry is directly
related to the quark transversity distribution $h_1(x)$, without
any additional terms.

{\it Approach 2} is to assume that the interaction dependent
twist-3 part, $\tilde{h}_L(x)$, is zero, thus we can also assume
that $\tilde{H}(z)$ is zero \cite{Kot99}. Then by neglecting the
term proportional to the current quark mass, one can obtain a
Wandzura-Wilczek type relation \cite{jaffe92,tan94}
\begin{equation}
h_{1L}^{\perp (1) }(x)=-x^2 \int_x^1 \d \xi
\frac{h_1(\xi)}{\xi^2}.
\end{equation}
It follows, from Eq.~(\ref{HLL}), that
\begin{equation}
h_{L}(x)=2x \int_x^1 \d \xi \frac{h_1(\xi)}{\xi^2}.
\end{equation}

Now we need the distribution functions $f_1(x)$ and $h_1(x)$ of
the target, and the fragmentation functions $D_1(z)$ and
$H_1^{\perp (1)}(z)$ for the produced meson. The usual
distribution functions $f_1(x)$ have been known with rather high
precision, and we adopt two model parametrizations : a
quark-diquark model and a pQCD based analysis \cite{MSY10}. The
transversity distributions $h_1(x)$ have not been measured yet,
but can be roughly predicted, and we adopt those given in the two
models, with different flavor and spin structure \cite{MSY10}.
Please see Ref.~\cite{MSY10} for detailed descriptions and
references. For the fragmentation functions of the pion, we adopt
the new parametrization presented in \cite{Kre01}, with a complete
set of both favored and unfavored fragmentation functions.

The so called Collins fragmentation function $H_1^{\perp (1)}(z)$,
which describes the transition of a transversely polarized quark
into a pion, has not been systematically measured yet, and it is
also theoretically less known. There has been a so called Collins
parametrization \cite{Col93} of this fragmentation function,
\begin{equation}
A_C(z,k_{\perp})=\frac{|k_{\perp}| H_1^{\perp
q}(z,z^2k^2_{\perp})}{M_h D_1^q(z,z^2k^2_{\perp})} =\frac{M_C
|k_{\perp}|}{M_C^2+|k_{\perp}^2|}, \label{CollinsAC}
\end{equation}
with $M_C$ being a typical hadronic scale around $0.3 \to 1$ GeV.
Assuming a Gaussian type of the quark transverse momentum
dependence in the unpolarized fragmentation function
\begin{equation}
D_1^q(z,z^2k^2_{\perp})=D_1^q(z) \frac{R^2}{\pi z^2} \exp(-R^2
k_{\perp}^2),
\end{equation}
one obtains
\begin{equation}
H_1^{\perp (1) q}(z)=D_1^q(z) \frac{M_C}{2M_h} \left(1-M_C^2 R^2
\int_0^{\infty} \d x \frac{\exp(-x)}{x+M_C^2 R^2}\right),
\label{CollinsH1}
\end{equation}
where $R^2=z^2/\left<P^2_{h\perp}\right>$, an
$\left<P^2_{h\perp}\right>=z^2\left<k_{\perp}^2\right>$ is the
mean-square momentum that the hadron acquires in the quark
fragmentation process. We will set the parameters $M_C=0.7$ GeV
and $\left<P^2_{h\perp}\right>=(0.44)^2$ GeV$^2$ as they are
consistent with the spin asymmetry measured at HERMES
\cite{Kor00}. It has been recently indicated by Efremov, Goeke,
and Schweitzer \cite{EGS}, that the uncertainties related to the
magnitude of the Collins fragmentation functions are still rather
big, and can vary within a factor of 2. So we consider a further
option for Approach 1 and Approach 2 by simply multiplying by a
factor 2 to the Collins parametrization (\ref{CollinsAC}) and
(\ref{CollinsH1}).

\vspace{0.3cm}
\begin{figure}[htb]
\begin{center}
\leavevmode {\epsfysize=7.5cm \epsffile{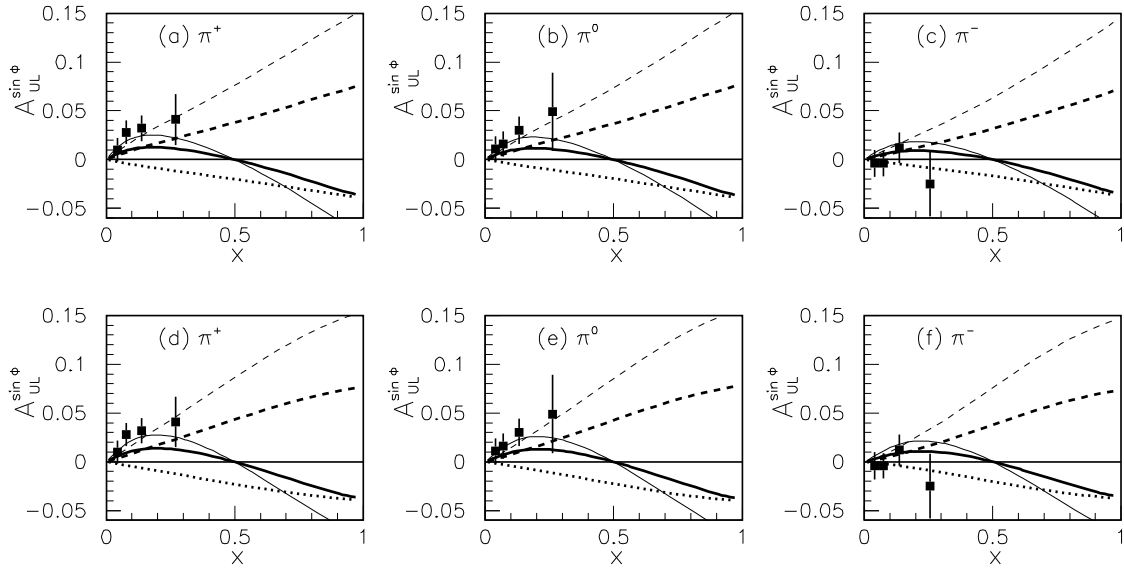}}
\end{center}
\caption[*]{\baselineskip 13pt The azimuthal asymmetries
$A_{UL}^{\sin \phi}$ for semi-inclusive pion production in deep
inelastic scattering of unpolarized charged lepton on the
longitudinally polarized proton target, with polarization $S=0.86$
\cite{Kor00}. The upper row corresponds to (a) $\pi^{+}$, (b)
$\pi^{0}$, and (c) $\pi^-$ production with the distribution
functions in the quark diquark model, and the lower row
corresponds to (d) $\pi^{+}$, (e) $\pi^{0}$, and (f) $\pi^-$
production with the distribution functions in the pQCD based
analysis. The thick dotted, dashed, and solid curves correspond to
the calculated results for Leading Approach, Approach 1, and
Approach 2, and the thin dashed and solid curves correspond to the
calculated results for Approach 1 and Approach 2 with an
additional factor 2 in the Collins fragmentation functions. Both
the favored and unfavored fragmentation functions for the pions
are included in the calculation.}\label{msy13f1}
\end{figure}

\vspace{0.3cm}
\begin{figure}
\begin{center}
\leavevmode {\epsfysize=7.5cm \epsffile{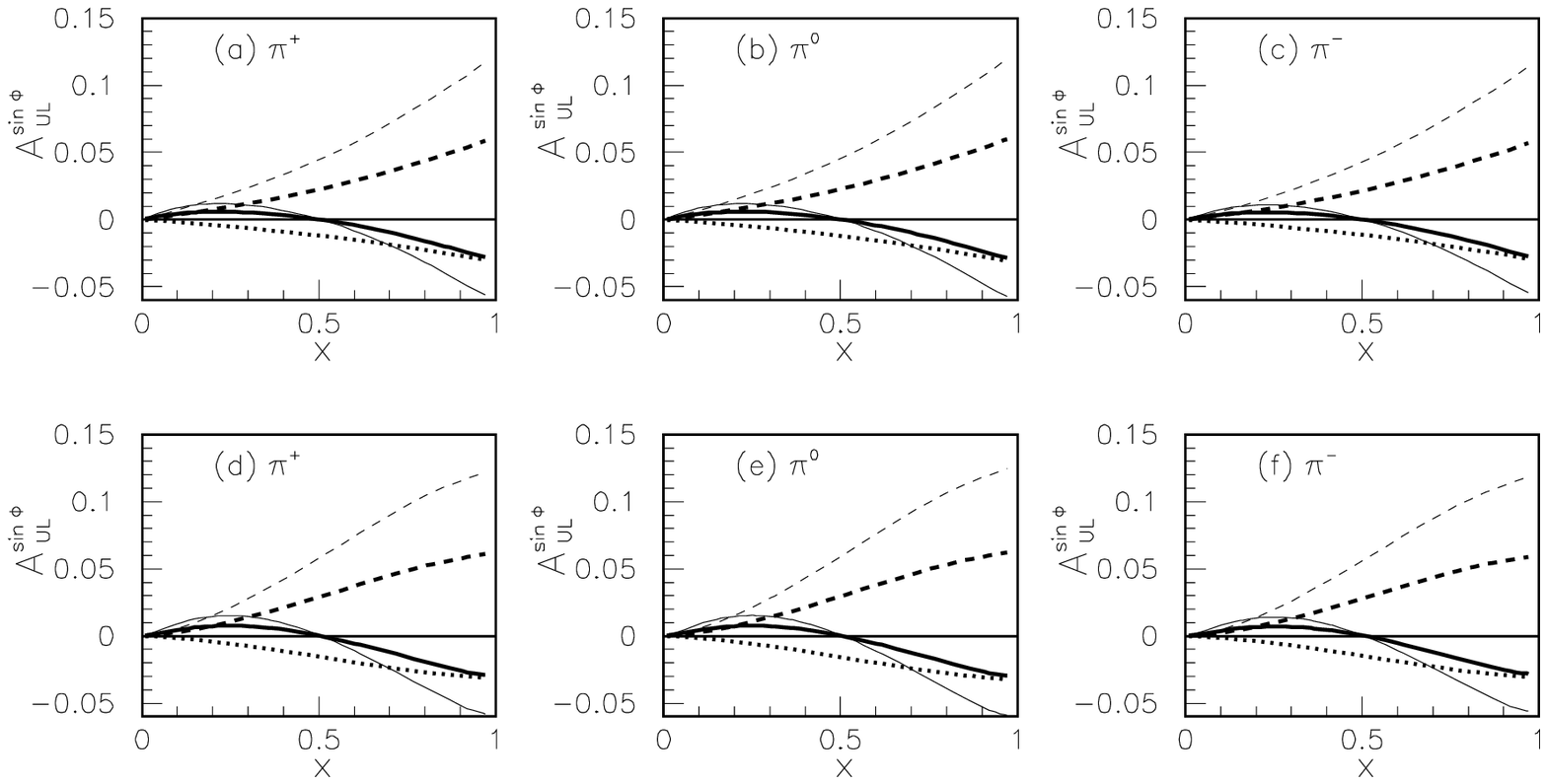}}
\end{center}
\caption[*]{\baselineskip 13pt The same as Fig.~\ref{msy13f1}, but
for the {\it deuteron} target with polarization $S=0.75$.
}\label{msy13f2}
\end{figure}

\vspace{0.3cm}
\begin{figure}
\begin{center}
\leavevmode {\epsfysize=7.5cm \epsffile{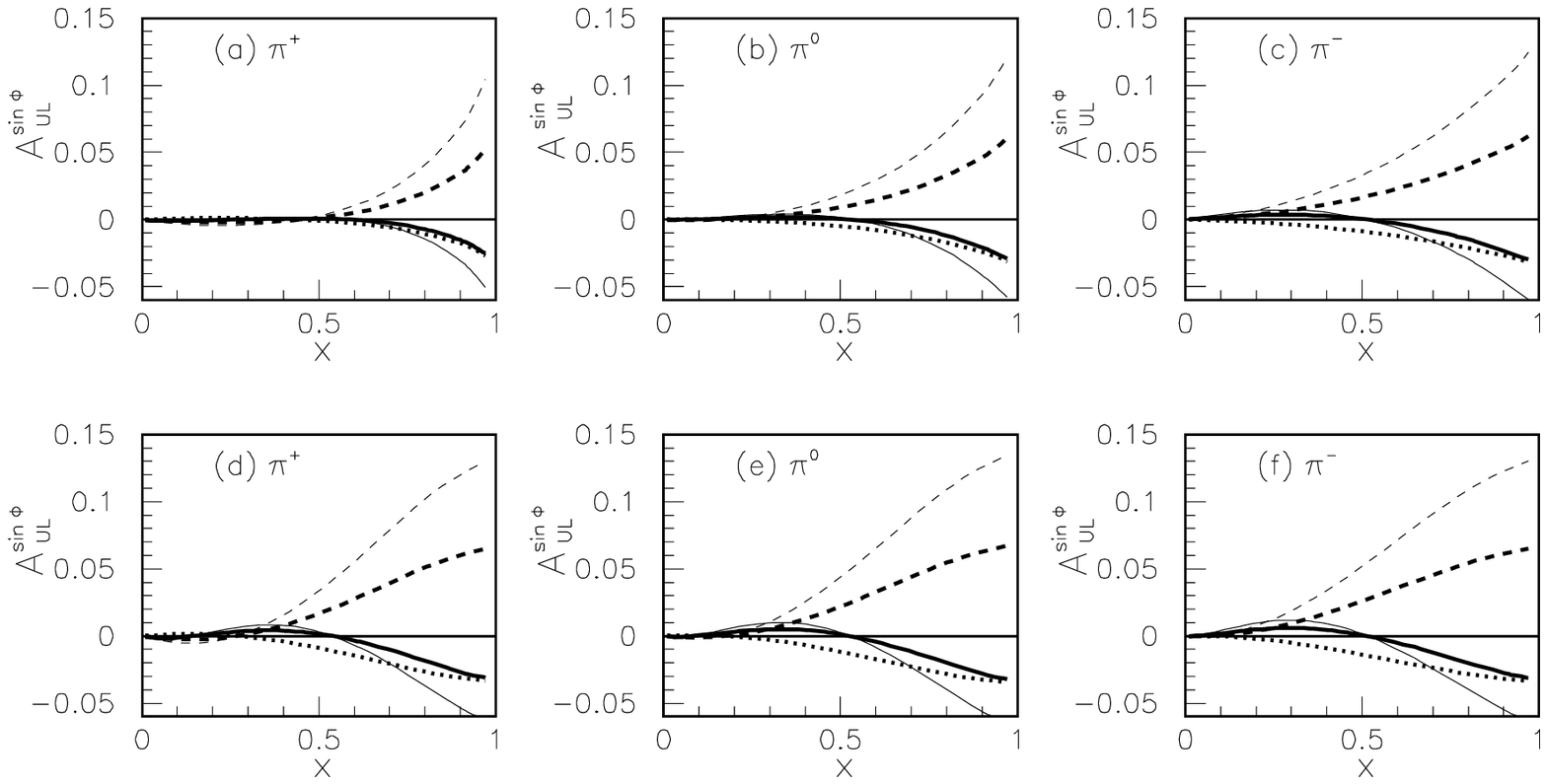}}
\end{center}
\caption[*]{\baselineskip 13pt The same as Fig.~\ref{msy13f1}, but
for the {\it neutron} target with polarization $S=0.75$.
}\label{msy13f3}
\end{figure}

\vspace{0.3cm}
\begin{figure}
\begin{center}
\leavevmode {\epsfysize=7.5cm \epsffile{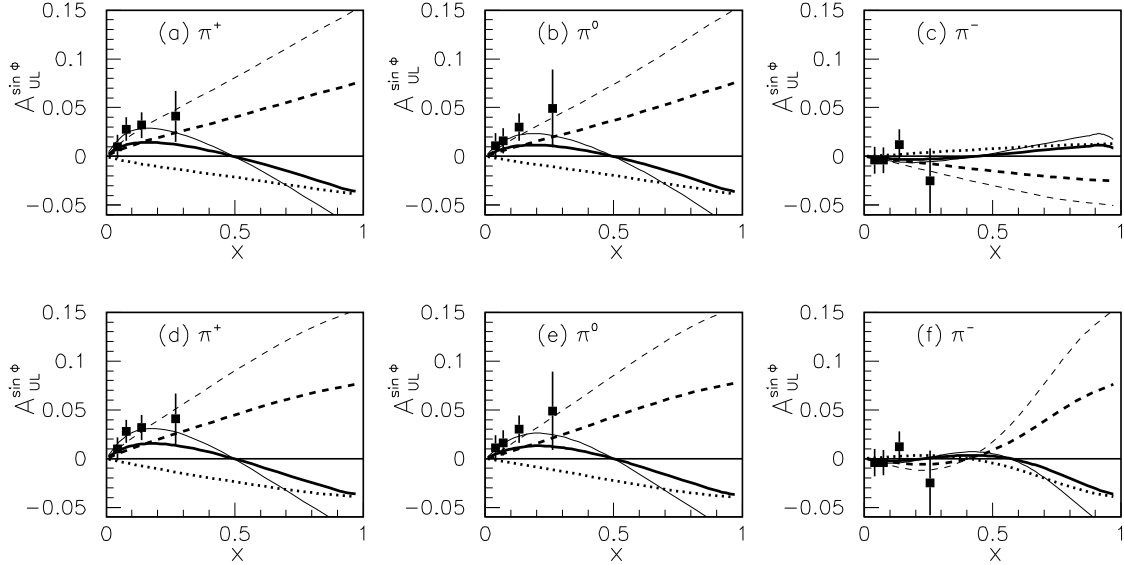}}
\end{center}
\caption[*]{\baselineskip 13pt The same as Fig.~\ref{msy13f1}, but
with only favored fragmentation functions for the pions being
included. }\label{msy13f4}
\end{figure}

In Figs.~\ref{msy13f1}-\ref{msy13f3}, we present predictions for
the azimuthal asymmetries $A_{UL}^{\sin \phi}$ for pion production
in proton, deuteron, and neutron targets respectively. It can be
seen that the predictions are significantly different for the five
cases: Leading Approach, Approach 1, Approach 2, and two further
options for Approach 1 and Approach 2 with the Collins
parametrization multiplied by a factor 2. By comparison with the
experimental data, we find that the Leading Approach seems to have
opposite trend with the data, except that for $\pi^{-}$. This
means that the available data on the azimuthal asymmetries by
HERMES collaboration cannot be interpreted as a direct measurement
of the transversity distributions of the nucleon, and/or the $1/Q$
term is not negligible at the HERMES energy. Predictions of
Approach 1 and Approach 2 can be compatible with the available
experimental data by including the uncertainties in the Collins
parametrization of T-odd fragmentation functions, but the two
approaches differ significantly at large $x$. Both the quark
diquark model and the pQCD base analysis give similar predictions
for the proton target. So the HERMES observation of the azimuthal
asymmetries may still provide information on the transversity
distributions of the nucleon at small $x$, as well as on the
Collins fragmentation functions. To check the effects from the
unfavored fragmentation functions, we also present in
Fig.~(\ref{msy13f4}) the predictions on the azimuthal asymmetries
with only favored fragmentation functions included. We find big
difference for $\pi^-$ production by comparing
Fig.~(\ref{msy13f4}) with Fig.~(\ref{msy13f1}). This implies that
the unfavored fragmentation functions are important for $\pi^-$
production from a proton target, as the $u$ quark content is
dominant at large $x$ in the target, and consequently, the
unfavored $u \to \pi^-$ fragmentation is important. Further
theoretical and experimental studies are still needed to
discriminate between different distribution functions and
fragmentation functions as reflected in Approach 1 and Approach 2.

We also mention here that an alternative mechanism for the
azimuthal asymmetries has been proposed
\cite{BHS,Collins,Ji:2002aa}. Here it was shown that the QCD
final-state interactions (gluon exchange) between the struck quark
and the proton spectators in semi-inclusive deep inelastic lepton
scattering can produce single-spin asymmetries which survive in
the Bjorken limit. This provides a physical explanation, within
QCD, of these asymmetries, and it also predicts that the
initial-state interactions from gluon exchange between the
incoming quark and the target spectator system lead to
leading-twist single-spin asymmetries in the Drell-Yan process
$H_1 H_2 \to \ell^+ \ell^- X$ \cite{Collins,BHS2}.
So our study suggests that we also need to consider the new
mechanism as a plausible source for the azimuthal asymmetries
observed by HERMES.

\begin{figure}[htb]
\begin{center}
\leavevmode {\epsfysize=7.5cm \epsffile{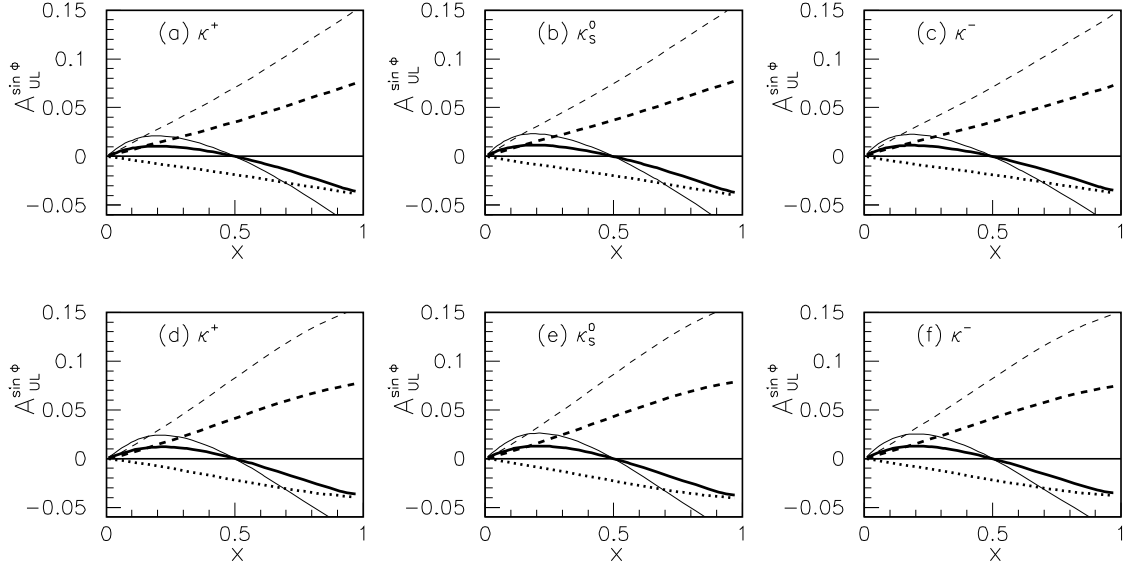}}
\end{center}
\caption[*]{\baselineskip 13pt The azimuthal asymmetries
$A_{UL}^{\sin \phi}$ for semi-inclusive kaon production in deep
inelastic scattering of unpolarized charged lepton on the
longitudinally polarized proton target, with polarization
$S=0.86$. The upper row corresponds to (a) $K^{+}$, (b) $K^{0}_S$,
and (c) $K^-$ production with the distribution functions in the
quark diquark model, and the lower row corresponds to (d) $K^{+}$,
(e) $K^{0}_S$, and (f) $K^-$ production with the distribution
functions in the pQCD based analysis. The thick dotted, dashed,
and solid curves correspond to the calculated results for Leading
Approach, Approach 1, and Approach 2, and the thin dashed and
solid curves correspond to the calculated results for Approach 1
and Approach 2 with an additional factor 2 in the Collins
fragmentation functions. Both the favored and unfavored
fragmentation functions for the kaons are included in the
calculation.}\label{msy13f5}
\end{figure}

\begin{figure}
\begin{center}
\leavevmode {\epsfysize=7.5cm \epsffile{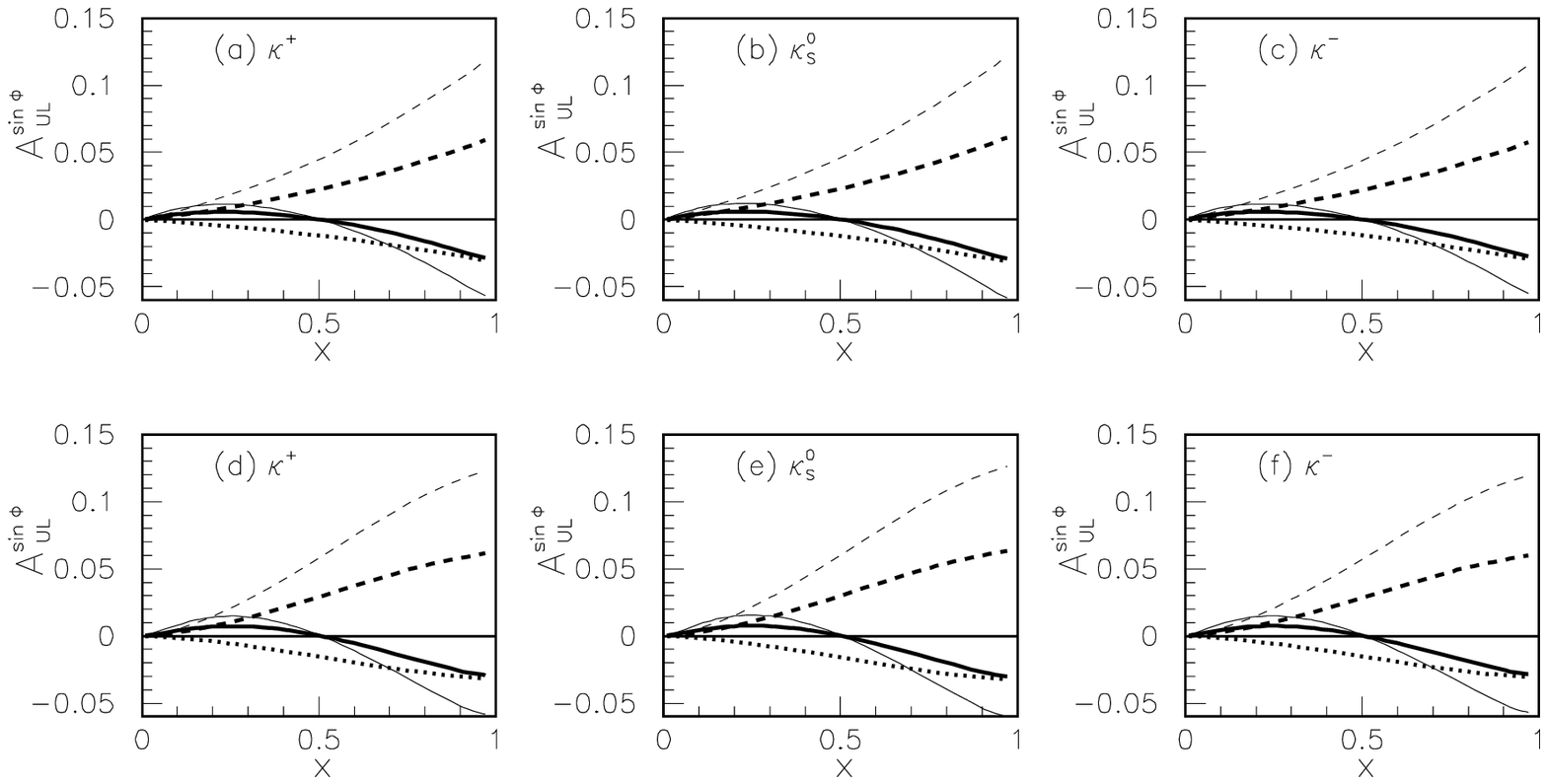}}
\end{center}
\caption[*]{\baselineskip 13pt The same as Fig.~\ref{msy13f5}, but
for the {\it deuteron} target with polarization $S=0.75$.
}\label{msy13f6}
\end{figure}

\begin{figure}
\begin{center}
\leavevmode {\epsfysize=7.5cm \epsffile{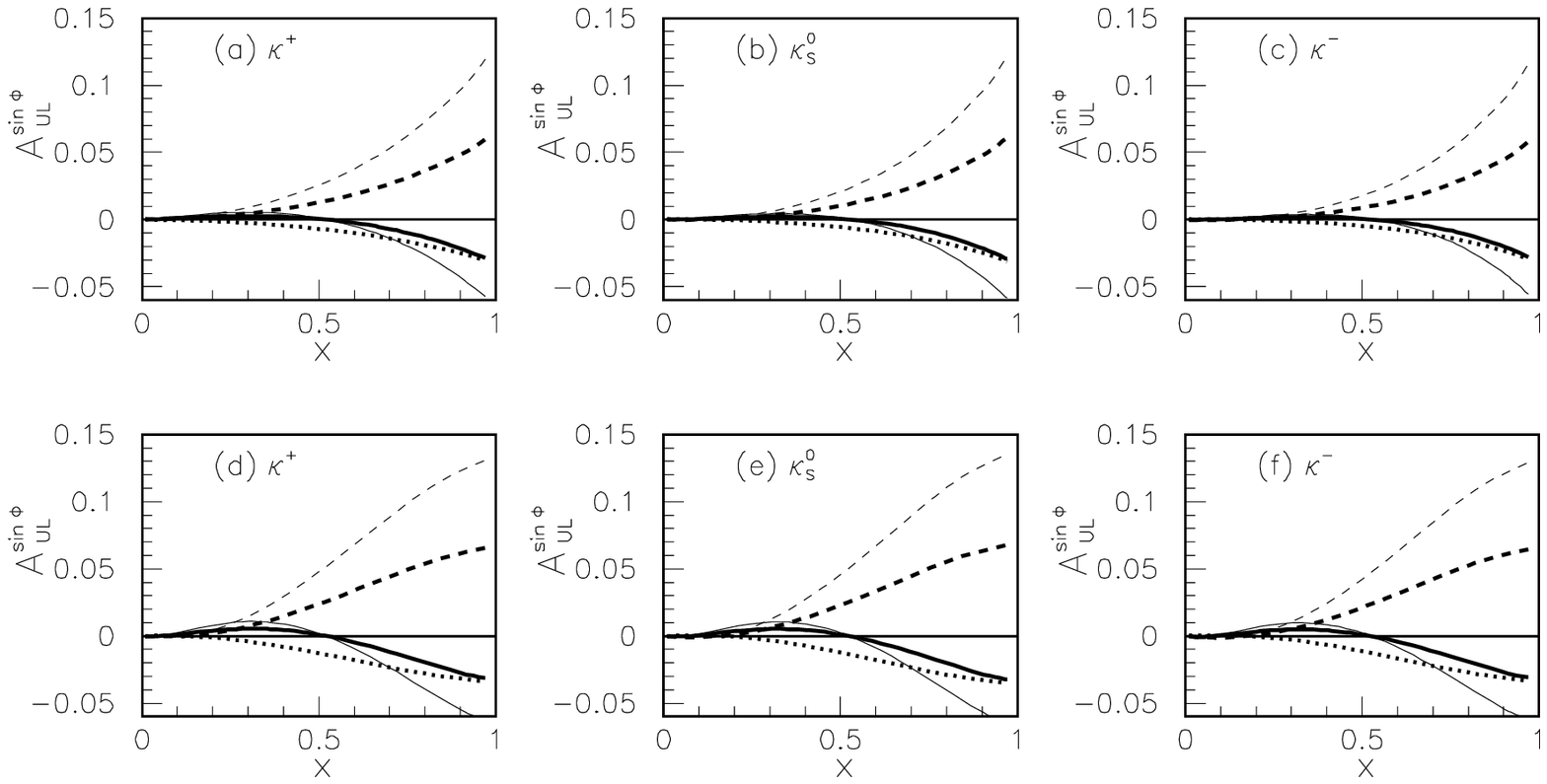}}
\end{center}
\caption[*]{\baselineskip 13pt The same as Fig.~\ref{msy13f5}, but
for the {\it neutron} target with polarization $S=0.75$.
}\label{msy13f7}
\end{figure}

\begin{figure}
\begin{center}
\leavevmode {\epsfysize=7.5cm \epsffile{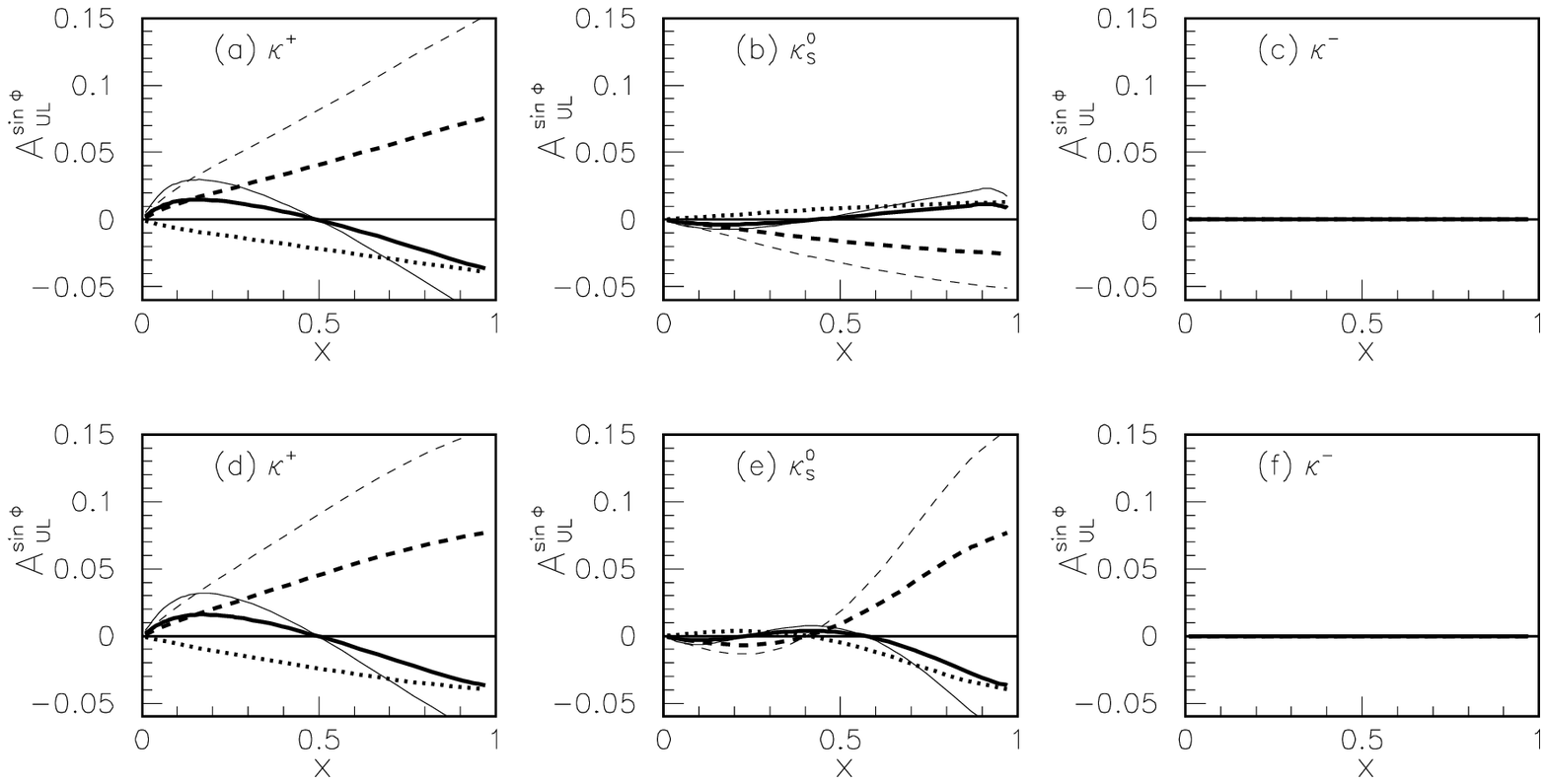}}
\end{center}
\caption[*]{\baselineskip 13pt The same as Fig.~\ref{msy13f5}, but
with only favored fragmentation functions for the kaons being
included. }\label{msy13f8}
\end{figure}

The HERMES collaboration will also measure the azimuthal
asymmetries for the kaon production. Thus it is necessary to make
predictions of azimuthal asymmetries $A_{UL}^{\sin \phi}$ for
$K^+$, $K^{0}_S$, and $K^-$ production from proton, deuteron, and
neutron targets respectively. We present our numerical results
with different approaches and options for the distribution
functions and fragmentation functions in
Figs.~\ref{msy13f5}-\ref{msy13f7}. In principle, the flavor
structure of the kaon fragmentation functions is more complicated
than that of the pion, but the available parametrizations
\cite{BKKK,BKKK0,DK1} only make distinction between the favored
fragmentation functions $D^{K}$, which are related to the valence
quarks of the kaon, and the unfavored fragmentation functions
$\hat{D}^{K}$, which are related to the light-flavor sea quarks of
the kaon. For $K^{\pm}$ we have \cite{BKKK}
\begin{equation}
\begin{array}{ll}
D^{K^{\pm}}(z)=0.31 z^{-0.98}(1-z)^{0.97},\\
\hat{D}^{K^{\pm}}(z)=1.08 z^{-0.82}(1-z)^{2.55}, \label{DK}
\end{array}
\end{equation}
and for $K_S^{0}$ we have \cite{BKKK0}
\begin{equation}
\begin{array}{ll}
D^{K^{0}_{S}}(z)=0.53 z^{-0.57}(1-z)^{1.87},\\
\hat{D}^{K^0_{S}}(z)=1.45 z^{-0.62}(1-z)^{3.84}. \label{DK0}
\end{array}
\end{equation}
The predictions given in Fig.~\ref{msy13f5}-\ref{msy13f7} are with
both favored and unfavored fragmentation functions included.  To
reflect the influence from the unfavored fragmentation functions
for kaon production, we give in Fig.~\ref{msy13f8} also
predictions for the azimuthal asymmetries of kaon production, for
a proton target, and with only favored fragmentation functions
$D^{K}$ included. We find, by comparing Fig.~\ref{msy13f5} with
Fig.~\ref{msy13f8}, that the unfavored fragmentation functions
also play an important role in $K^{-}$ and $K^0_S$ production.
This is easy to understand, since the favored fragmentation should
be $\bar{u} \to K^-$ and $s \to K^-$ for $K^-$ production, and
both the $\bar{u}$ and $s$ quarks belong to the sea content in the
nucleon target. The sea in both the quark diquark model and pQCD
based analysis is assumed to be unpolarized. Thus the azimuthal
asymmetries for $K^-$ production should have zero values in the
two models with only favored fragmentation functions included. But
the situation will be different if both favored and unfavored
fragmentation functions are included, as the unfavored $u$
fragmentation will be enhanced due to the dominant $u$ quark
content at large $x$ in the proton. The valence $u$ quark is
positively polarized in both the quark diquark model and the pQCD
based analysis, thus the $K^-$ production is sensitive to the
unfavored fragmentation functions. Therefore we conclude that the
unfavored fragmentation functions cannot be neglected for $K^-$
and $K^0_S$ production in semi-inclusive DIS process, and this is
different from the predictions in \cite{EGS}.

\vspace{0.3cm}

\vspace{0.3cm}

\vspace{0.3cm}

\vspace{0.3cm}

In summary, we reanalyzed the azimuthal spin asymmetries for pion
production in deep inelastic scattering of unpolarized charged
lepton beam on the longitudinally polarized nucleon target, by
taking into account an important sign correction to previous
formulas. We find that this sign correction causes significant
changes to the predictions, and we also find that the predictions
are very different with different approaches for distribution
functions and fragmentation functions. With the result that now it
is difficult to use this process for a direct measurement of
transversity at large $x$. As a results of similarity between the
results of Approach 1 and Approach 2 at small $x$, one can still
attribute the recent HERMES measurements as a rough estimate of
the transversity distributions of the nucleon at small $x$, thus
the HERMES data can be used to provide some useful constraints on
the transversity distributions and on the T-odd Collins
fragmentation functions. Further theoretical and experimental
studies are still necessary to reveal various distribution
functions and fragmentation functions. This includes taking into
account new physical mechanisms in these processes
\cite{BHS,Collins,Ji:2002aa,BHS2}.
Predictions for the azimuthal spin asymmetries of kaon productions
are also presented with different approaches and options for the
distribution functions and fragmentation functions, and we
conclude that the unfavored fragmentation functions cannot be
neglected for $K^-$ and $K^0_S$ productions in semi-inclusive deep
inelastic scattering.

{\bf Acknowledgments: } This work is partially supported by
National Natural Science Foundation of China under Grant Numbers
19975052, 10025523, 90103007 and 10175074, by Fondecyt (Chile)
3990048 and 8000017, and also by Foundation for University Key
Teacher by the Ministry of Education (China). We are grateful to
A.V.~Efremov, K.~Goeke, and P.~Schweitzer for the correspondences
concerning the correction. We also thank W.-D. Nowak for helpful
discussions and suggestions.


\end{document}